\documentclass{revtex4}
\usepackage{amssymb}
\usepackage{amsmath}

\input{tcilatex}
\begin{document}

\title{Quantum tunneling from rotating black holes with scalar hair in three
dimensions}
\author{I. Sakalli$^{\ast }$}
\author{H. Gursel$^{\dag }$}
\affiliation{Department of Physics, Eastern Mediterranean University,}
\affiliation{G. Magusa, North Cyprus, Mersin-10, Turkey.}
\affiliation{$^{\ast }$izzet.sakalli@emu.edu.tr}
\affiliation{$^{\dagger }$huriye.gursel@cc.emu.edu.tr}

\begin{abstract}
We study the Hawking radiation (HR) of scalar and Dirac particles (fermions)
emitted from a rotating scalar hair black hole (RSHBH) within the context of
three dimensional ($3D$) Einstein gravity using non-minimally coupled scalar
field theory. Amalgamating the quantum tunneling approach with the
Wentzel--Kramers--Brillouin (WKB) approximation, we obtain the tunneling
rates of the outgoing particles across the event horizon. Inserting the
resultant tunneling rates into the Boltzmann formula, we then obtain the
Hawking temperature ($T_{H}$) of the $3D$ RSHBH.
\end{abstract}

\maketitle

\section{Introduction}

The most significant prediction of Einstein's field equations is the
existence of black holes (BHs) \cite{Gravitation}. A BH is a region of
spacetime where the gravity is sufficiently strong to trap light. According
to general relativity (GR), classically, BHs are perfect absorbers from
which emission is impossible. However, this idea was dramatically overturned
by quantum mechanics (QM). In a remarkable discovery, Hawking \cite{Hawking}
demonstrated the mechanism of radiation by BHs. He showed that, rather than
being completely \textquotedblleft black\textquotedblright\ as predicted by
GR, BHs emit thermal radiation; the so-called HR. His discovery can be
rederived by various methods (see for instance \cite{Wald1,Vanzo} and
references therein), leading to quantum gravity (QG), an intriguing theory
that relates GR and QM \cite{Rovelli}. On the other hand, this revolutionary
result is not fully compatible with the principles of QM because thermal
radiation provides no information about the object that sourced the BH.
Consequently, once the BH has evaporated this information, it is lost
forever, violating a basic tenet of QM; that information must be conserved.
This contradiction is known as the information loss paradox (ILP) \cite%
{Hawking2}. ILP has been extensively referenced in \cite{InfoBH}. Moreover,
HR can be considered as the quantum tunneling of particles from BH horizons 
\cite{Kraus,Parikh,Iso}. According to this theorem, when a virtual particle
pair is created just outside the BH horizon, the antiparticle (negative
energy particle) can tunnel through the BH horizon by a process similar to
QM tunneling, whereas the real particle (positive energy particle) is
ejected into spatial infinity. Inversely, by particle--antiparticle
symmetry, a virtual pair can be created just inside the horizon. In this
case, the real particle can tunnel inward, while the antiparticle remains
inside the BH \cite{Izzet}.

Traditionally, BHs are regarded as very simple objects that can be
completely characterized by 3 parameters: mass, charge, and angular
momentum. Because of their simplicity, Wheeler \cite{Gravitation}, who named
BHs, insisted that \textquotedblleft BHs have no hair.\textquotedblright\ In
honor of this statement, the traditional concept of BHs is known as the
no-hair theorem (NHT) \cite{NoHair}. However, some researchers have
suggested that BHs might be hairier than previously thought. In anew
mechanism developed by Herdeiro and Radu \cite{Herdeiro,Herdeiro2}, scalar
and other types of fields (in principle) admit hairy BHs. Explicit solutions
of hairy BHs have been reported in the literature (e.g., \cite%
{Herdeiro,Herdeiro2,Herdeiro3,SH1,SH2,SH3,SH4} and references therein).
Cosmologists frequently use scalar fields to model the evolution of the
Universe \cite{Zuckerman}. However, the physical properties of scalar hair
BHs, and their role in the natural cosmos, require further study. On the
other hand, since the seminal works of Deser et al. \cite%
{Deser1,Deser2,Deser3} and Witten \cite{Witten1,Witten2}, the foundations of
classical gravity and QG have increasingly been investigated by GR in $3D$
spacetime \cite{Carlip}. Therefore, we here consider a $3D$ RSHBH \cite%
{RHSBH1,RHSBH2,RHSBH3,RHSBH4,RHSBH5} as a solution to the Einstein gravity
equations with a non-minimally coupled scalar field $\phi $. In the absence
of the scalar field ($\phi =0$), the $3D$ RSHBH reduces to the well-known
rotating Ba\~{n}ados--Teitelboim--Zanelli (BTZ) BH \cite{BTZ1,BTZ2,Carlip2}.

To investigate the HR of a $3D$ RSHBH, we obtain the tunneling rate of the
outgoing scalar and spinor particles penetrating the event horizon of the $%
3D $ RSHBH. In the derivation, we combine the Hamilton--Jacobi (HJ) ans\"{a}%
tze with the WKB approximation \cite{WKB}. Inserting the computed tunneling
rates into the Boltzmann formula \cite{Kraus2}, we then prove that the
standard $T_{H}$ of the $3D$ RSHBH is obtainable for all particle types.

The remainder of this paper is organized as follows. Section 2 briefly
reviews the geometrical and physical properties of the $3D$ RSHBH. In Sec.
3, we explore the HR of scalar particles emitted from the $3D$ RSHBH using
the Klein--Gordon equation (KGE). Section 3 is devoted to the HR of fermions
tunneling from the $3D$ RSHBH. Conclusions are presented in Sec. 4.

Throughout the paper, we use units wherein $c=G=k_{B}=1$.

\section{Features of $3D$ RSHBH}

The general solution to the action of the $3D$ Einstein gravity with a
non-minimally coupled scalar field $\phi $, which describes rotating BHs
with scalar hair, was first reported by Xu and Zhao \cite{RHSBH1}. The
metric of the $3D$ RSHBH is given by

\begin{equation}
ds^{2}=fdt^{2}-\frac{dr^{2}}{f}-r^{2}(d\theta +\ f^{\theta }dt)^{2},
\label{1}
\end{equation}

with

\begin{equation}
f=\frac{r^{2}}{36l^{2}}\left( J^{2}l^{2}x^{2}-12Ml^{2}x+36\right) ,
\label{2}
\end{equation}

\begin{equation}
f^{\theta }=-\frac{Jx}{6},  \label{3n}
\end{equation}

where

\begin{equation}
x=\frac{3r+2B}{r^{3}}.  \label{4n}
\end{equation}

The parameters $M$ and $J$ denote the physical mass and angular momentum of
the BH, respectively. $B$ is an integration constant and $\Lambda =\frac{1}{%
l^{2}}$ is the cosmological constant. Without loss of generality, we assume
that $B$ is a positive real number. Meanwhile, it is worth noting that when $%
B=0$ metric (1) describes the rotating BTZ BH \cite{RHSBH3}.

Equation (2) can be rewritten as follows

\begin{equation}
f=(\frac{Jr}{6})^{2}\left[ (x-x_{1})(x-x_{2})\right] ,  \label{5nn}
\end{equation}

where%
\begin{equation}
x_{k}=\frac{6M}{J^{2}}-(-1)^{k}\frac{6}{J^{2}l}\sqrt{M^{2}l^{2}-J^{2}},\text{
\ \ \ \ \ \ \ }(k=1,2),  \label{6nn}
\end{equation}

with\ ($x_{1},x_{2}$) being real positive quantities. On the other hand, we
immediately observe that the $3D$ RSHBH is constrained by $Ml\geqslant J$.

We now investigate the location of the event horizon ($r_{h}$) of the $3D$
RSHBH. Since $f(r_{h})=0$, Eq. (4) gives

\begin{equation}
r_{h(k)}^{3}+\widetilde{A}_{k}r_{h(k)}+\widetilde{B}_{k}=0.  \label{cubic7}
\end{equation}

where

\begin{equation}
\widetilde{A}_{k}=-\frac{3}{x_{k}},  \label{8m}
\end{equation}

\begin{equation}
\widetilde{B}_{k}=-\frac{2B^{3}}{x_{k}}.  \label{9m}
\end{equation}

Equation (7) is merely a cubic equation \cite{CubicEq}, whose discriminant
is given by

\begin{equation}
D_{k}=\frac{4\widetilde{A}_{k}^{3}+27\widetilde{B}_{k}^{2}}{108}.
\label{10m}
\end{equation}

Substituting $\widetilde{A}_{k}$ and $\widetilde{B}_{k}$ into Eq. (10), we
obtain

\begin{equation}
D_{k}=B^{6}\frac{\left( x_{k}-1\right) }{x_{k}^{3}}.  \label{11m}
\end{equation}

If $D_{k}>0$ or $x_{k}>1,$ we have a single positive real root:%
\begin{equation}
r_{h(k)}=\left( -\frac{\widetilde{B}_{k}}{2}+\sqrt{D_{k}}\right)
^{1/3}+\left( -\frac{\widetilde{B}_{k}}{2}-\sqrt{D_{k}}\right) ^{1/3}.
\label{12m}
\end{equation}

Inserting Eqs. (9) and (11) into Eq. (12), we obtain

\begin{equation}
r_{h(k)}=\frac{B}{x_{k}}\left[ \left( x_{k}^{2}+\sqrt{x_{k}^{3}\left(
x_{k}-1\right) }\right) ^{1/3}+\left( x_{k}^{2}-\sqrt{x_{k}^{3}\left(
x_{k}-1\right) }\right) ^{1/3}\right] .  \label{13m}
\end{equation}

On the other hand, if $D_{k}<0$ or $0<x_{k}<1,$ we can define a new variable

\begin{equation}
\cos \alpha _{k}=\frac{r_{h(k)}}{2}\sqrt{-\frac{\widetilde{A}_{k}}{3}},
\label{14m}
\end{equation}

which transforms Eq. (7) into the following form

\begin{equation}
4\cos ^{3}\alpha _{k}-3\cos \alpha _{k}-\frac{3\widetilde{B}_{k}}{2%
\widetilde{A}_{k}}\sqrt{-\frac{3}{\widetilde{A}_{k}}}=0.  \label{15m}
\end{equation}

Recalling the identity

\begin{equation}
\cos \left( 3\theta \right) =4\cos ^{3}\theta +3\cos \theta ,  \label{16m}
\end{equation}

Eq. (15) becomes

\begin{equation}
\cos \left( 3\alpha _{k}\right) =\frac{3\widetilde{B}_{k}}{2\widetilde{A}_{k}%
}\sqrt{-\frac{3}{\widetilde{A}_{k}}.}  \label{17m}
\end{equation}

Thus, the solutions to Eq. (7) with $D_{k}<0$ are the roots

\begin{equation}
r_{h(k)}=\frac{2B}{\sqrt{x_{k}}}\cos \left[ \frac{1}{3}\cos ^{-1}\left( 
\sqrt{x_{k}}\right) -\frac{2\pi }{3}n\right] ,\text{ \ \ \ }(n=0,1,2,....).
\label{18m}
\end{equation}

The positive root solutions when $n=0$ contradict the result in \cite{RHSBH3}%
. Therefore, the event horizon reads

\begin{equation}
r_{h(k)}=\frac{2B}{\sqrt{x_{k}}}\cos \left[ \frac{1}{3}\cos ^{-1}\left( 
\sqrt{x_{k}}\right) \right] .  \label{19m}
\end{equation}

In \cite{RHSBH3,RHSBH4}, the mass ($M$), $T_{H}$, and BH\ entropy ($S_{BH}$)
of the $3D$ RSHBH are given by

\begin{equation}
M=\frac{J^{2}l^{2}(2B+3r_{h})^{2}+36r_{h}^{6}}{12l^{2}r_{h}^{3}(2B+3r_{h})},
\label{20}
\end{equation}

\begin{equation}
T_{H}=\frac{\hbar \kappa }{2\pi }=\frac{\hbar
(B+r_{h})[36r_{h}^{6}-J^{2}l^{2}(2B+3r_{h})^{2}]}{24\pi
l^{2}r_{h}^{5}(2B+3r_{h})},  \label{21n}
\end{equation}

\begin{equation}
S_{BH}=\frac{A_{H}}{4\hbar }[1-\frac{1}{8}\phi ^{2}(r_{h})]=\frac{4\pi
r_{h}^{2}}{B+r_{h}},  \label{22n}
\end{equation}

where $\phi (r)=\pm \sqrt{\frac{8B}{B+r}}$\ corresponds to the scalar field
in the $3D$ RSHBH spacetime and $\kappa $\ is the surface gravity \cite%
{Wald2} of the BH. The angular velocity of the $3D$ RSHBH is

\begin{equation}
\Omega _{H}=-\left. \frac{g_{t\theta }}{g_{\theta \theta }}\right\vert
_{r=rh}=-f^{\theta }(r_{h})=\frac{(3r_{h}+2B)J}{6r_{h}^{3}}.  \label{23n}
\end{equation}

$\Omega _{H}$ is precisely the angular rotation frequency of the BH. That
is, any test body dropped into the BH will circumnavigate the BH at this
frequency as it approaches the event horizon. Furthermore, one can easily
verify that the first law of thermodynamics \cite{RHSBH3} holds:

\begin{equation}
dM=T_{H}dS_{BH}+\Omega _{H}dJ.  \label{24}
\end{equation}

\section{Quantum Tunneling of Scalar Particles from 3D RSHBH}

In this section, we evaluate the $T_{H}$ of the $3D$ RSHBH via the
relativistic KGE for scalar particles. The massive KGE can be written as
follows:

\begin{equation}
\partial _{\mu }(\sqrt{-g}g^{\mu \nu }\partial _{\nu }\Psi _{0})+\frac{%
m_{0}^{2}}{\hbar ^{2}}\sqrt{-g}\Psi _{0}=0,  \label{25n}
\end{equation}

where $m_{0}$ denotes the mass of the scalar (spin--$0)$ particle, and $\Psi
_{0}$ represents the scalar field. Since there exist non-diagonal components
in the metric (1), Eq. (25) takes the following form

\begin{equation}
\partial _{t}\left[ rg^{tt}\partial _{t}\Psi _{0}\right] +\partial _{r}\left[
rg^{rr}\partial _{r}\Psi _{0}\right] +\partial _{\theta }\left[ rg^{\theta
\theta }\partial _{\theta }\Psi _{0}\right] +\partial _{t}\left[ rg^{t\theta
}\partial _{\theta }\Psi _{0}\right] +\partial _{\theta }\left[ rg^{\theta
t}\partial _{t}\Psi _{0}\right] -\frac{m_{0}^{2}}{\hbar ^{2}}r\Psi _{0}=0.
\label{26n}
\end{equation}

To apply the WKB approximation method, we assume an ans\"{a}tz of the form
(see, for instance, \cite{Darabi})

\begin{equation}
\Psi _{0}=\exp [\frac{i}{\hbar }I+I_{1}+O(\hbar )].  \label{27n}
\end{equation}

Taking leading powers of $\hbar ,$ in Eq. (26), we obtain

\begin{equation}
f^{-1}(\partial _{t}I)^{2}-f(\partial _{r}I)^{2}-\frac{f-r^{2}(f^{\theta
})^{2}}{fr^{2}}(\partial _{\theta }I)^{2}-2\frac{f^{\theta }}{f}(\partial
_{t}I\partial _{\theta }I)-m_{0}^{2}=0.  \label{28n}
\end{equation}

As spacetime is symmetric, we have the Killing vectors $\partial _{t}$\ and $%
\partial _{\theta }$.\ Thus, we can apply the separation of variables method
to the classical action $I(t,r,\theta )$:

\begin{equation}
I=-Et+L\theta +W(r)+c,  \label{29n}
\end{equation}

where $E$ and $L$ represent the energy and angular momentum of the scalar
particle, respectively, and $c$ is a complex constant. Using Eq. (29) in Eq.
(28), we obtain the following equation for $W(r)$:%
\begin{equation}
W_{\pm }(r)=\pm \dint \frac{\sqrt{(E+Lf^{\theta })^{2}-f\left[ \left( \frac{L%
}{r}\right) ^{2}+m_{0}^{2}\right] }}{f}.  \label{30n}
\end{equation}

Here, the positive and negative signs indicate that the scalar particles
move away from the event horizon (emission) and toward the event horizon
(absorption), respectively. On the other hand, since $f(r_{h})=0$, Eq. (30)
possesses a simple pole at $r=r_{h}$. Thus, the integral (30) can be solved
by the residue theorem. For this purpose, we expand the metric function $f$
in a Taylor series about $r_{h}$:

\begin{eqnarray}
f(r_{h}) &=&f(r_{h})+f^{\prime }(r_{h})(r-r_{h})+O[(r-r_{h})^{2}],  \notag \\
&\simeq &f^{\prime }(r_{h})(r-r_{h}).  \label{31n}
\end{eqnarray}%
\bigskip Here, prime "$\prime $" over a quantity denotes a derivative with
respect to $r$. Hence, Eq. (30) can be approximated as

\begin{equation}
W_{\pm }(r)=\pm \dint \frac{\widetilde{E}}{f^{\prime }(r_{h})(r-r_{h})},
\label{32n}
\end{equation}

where the modified energy $\widetilde{E}$ is given by

\begin{equation}
\widetilde{E}=E+f^{\theta }(r_{h})L=E-L\Omega _{H}.  \label{33n}
\end{equation}%
Integrating Eq. (32) with respect to $r$ (using the residue theorem for semi
circles), we obtain

\begin{equation}
W_{\pm }=\pm i\pi \frac{\widetilde{E}}{f^{\prime }(r_{h})}.  \label{34n}
\end{equation}

The probabilities of the particles entering and leaving the BH through the
event horizon, respectively, are given by

\begin{equation}
\Gamma _{absorption}=\exp (-\frac{2}{\hbar }\func{Im}I)=\exp [-\frac{2}{%
\hbar }\left( \func{Im}W_{-}+\func{Im}c\right) ],  \label{35n}
\end{equation}

\begin{equation}
\Gamma _{emission}=\exp (-\frac{2}{\hbar }\func{Im}I)=\exp [-\frac{2}{\hbar }%
\left( \func{Im}W_{+}+\func{Im}c\right) ].  \label{36n}
\end{equation}

Since the objects close to the event horizon are destined to be swallowed by
the BH, the absorption probability ($\Gamma _{absorption}$) should be
normalized to unity by choosing the imaginary part of the constant as $\func{%
Im}c=-\func{Im}W_{-}.$ As is already known, $\func{Im}W_{+}=-\func{Im}W_{-}$%
; consequently we have $\func{Im}c=\func{Im}W_{+}$. Therefore, the tunneling
rate of scalar particles escaping the event horizon of the $3D$ RSHBH from
the interior is given by

\begin{eqnarray}
\Gamma _{_{emission}} &=&\exp \left( -\frac{4}{\hbar }\func{Im}W_{+}\right) ,
\notag \\
&=&\exp \left( -\frac{4\pi \widetilde{E}}{\hbar f^{\prime }(r_{h})}\right) .
\label{37n}
\end{eqnarray}

Equation (37) can be explicitly rewritten as

\begin{equation}
\Gamma _{_{emission}}=\exp \left\{ -\frac{24\pi l^{2}r_{h}^{5}(2B+3r_{h})%
\widetilde{E}}{\hbar (B+r_{h})[36r_{h}^{6}-J^{2}l^{2}(2B+3r_{h})^{2}]}%
\right\} .  \label{38n}
\end{equation}

Recalling the Boltzmann factor \cite{Srinivasan}:

\begin{equation}
\Gamma =\exp (-\beta \omega ),  \label{39n}
\end{equation}

where $\beta $ and $\omega $ denote the inverse temperature and energy,
respectively; the surface temperature is calculated as

\begin{equation}
T=\frac{\hbar (B+r_{h})[36r_{h}^{6}-J^{2}l^{2}(2B+3r_{h})^{2}]}{24\pi
l^{2}r_{h}^{5}(2B+3r_{h})}.  \label{40n}
\end{equation}

This result is obviously consistent with Eq. (21). Consequently, we have
proven that the standard $T_{H}$ of the $3D$ RSHBH is recovered by the
scalar particles tunneling the event horizon.

\section{Quantum Tunneling of Dirac Particles From 3D RSHBH}

In this section, we evaluate the contribution of fermions to the HR of the $%
3D$ RSHBH using the uncharged Dirac equation (UDE). Spinors in $3D$
spacetime possess two components, corresponding to the positive and negative
energy eigenstates. Therefore, the UDE comprises a pair of coupled partial
differential equations. As demonstrated by Sucu and Unal \cite{Sucu}, in
flat spacetime we can apply the following constant Dirac matrices $\overline{%
\sigma }^{(a)}$ \cite{Pitelli,Gurtug}:

\begin{equation}
\overline{\sigma }^{(a)}=\left( \overline{\sigma }^{(0)},\overline{\sigma }%
^{(1)},\overline{\sigma }^{(2)}\right) ,  \label{41n}
\end{equation}

with

\begin{equation}
\overline{\sigma }^{(0)}=\sigma ^{(3)},\text{ \ \ \ \ \ \ \ \ }\overline{%
\sigma }^{(0)}=i\sigma ^{(1)},\text{ \ \ \ \ \ \ \ \ \ \ }\overline{\sigma }%
^{(2)}=i\sigma ^{(2)},  \label{42n}
\end{equation}

where $\sigma ^{(1)},\sigma ^{(2)},\sigma ^{(3)}$ are the well-known Pauli
matrices. The $\overline{\sigma }^{(a)}$'s satisfy the following
anticommutation relation:

\begin{equation}
\overline{\sigma }^{(a)}\overline{\sigma }^{(b)}+\overline{\sigma }^{(a)}%
\overline{\sigma }^{(b)}=2\eta ^{\left( ab\right) },  \label{43n}
\end{equation}

where $\eta ^{\left( ab\right) }$ denotes the metric of the $3D$ Minkowski
spacetime. Using the triad of components $e_{\left( a\right) }^{\mu }$
composing the orthonormal frame, we can obtain the curved spacetime
dependent matrices $\overline{\sigma }^{\mu }$ in terms of the constant
matrices as follows:

\begin{equation}
\overline{\sigma }^{\mu }=e_{\left( a\right) }^{\mu }\overline{\sigma }%
^{(a)}.  \label{44n}
\end{equation}

The Greek indices ($\mu ,\nu $) represent the external (global) spacetime
indices, and the Latin indices ($a,b$) denote the internal (local) indices.
Hence, the metric tensor is given by

\begin{equation}
g_{\mu \nu }=e_{\mu }^{\left( a\right) }e_{\nu }^{\left( b\right) }\eta
_{\left( ab\right) }.  \label{45n}
\end{equation}

Ultimately, as formulated in \cite{Sucu}, the UDE of a fermion (spin-$\frac{1%
}{2}$) with mass $m_{s}$ and wave function (spinor) $\Psi _{s}$ in $3D$
curved spacetime is given by 
\begin{equation}
i\overline{\sigma }^{\mu }\left[ \partial _{\mu }-\Gamma _{\mu }\right] \Psi
_{s}=\frac{m_{s}}{\hbar }\Psi _{s},  \label{46n}
\end{equation}%
where $\Gamma _{\mu }$\ is the spinorial affine connection:

\begin{equation}
\Gamma _{\mu }=H_{\lambda \nu \mu }s^{\lambda \nu }.  \label{47n}
\end{equation}

The rank-$3$ tensor $H_{\lambda \nu \mu }$ and the spin operator $s^{\lambda
\nu }$ are, respectively, given by

\begin{equation}
H_{\lambda \nu \mu }=\frac{1}{4}g_{\lambda \alpha }\left[ e_{\nu ,\mu
}^{\left( i\right) }e_{\left( i\right) }^{\alpha }-\Gamma _{\nu \mu
}^{\alpha }\right] ,  \label{48n}
\end{equation}

\begin{equation}
s^{\lambda \nu }=\frac{1}{2}\left[ \sigma ^{\lambda },\sigma ^{\nu }\right] ,
\label{49n}
\end{equation}

where $\Gamma _{\nu \mu }^{\alpha }$ is the Christoffel symbol. The
following is a possible triad for metric (1)

\begin{equation}
e_{\mu }^{(i)}=%
\begin{pmatrix}
-\sqrt{f} & 0 & -f^{\theta }r \\ 
0 & -\frac{1}{\sqrt{f}} & 0 \\ 
0 & 0 & -r%
\end{pmatrix}%
,  \label{50n}
\end{equation}

which yields the following constant matrices:

\begin{equation}
\overline{\sigma }^{\mu }=\left( -\frac{\sigma ^{(3)}}{\sqrt{f}},-i\sqrt{f}%
\sigma ^{(1)},\frac{rf^{\theta }\sigma ^{(3)}-i\sqrt{f}\sigma ^{(2)}}{\sqrt{f%
}r}\right) .  \label{51n}
\end{equation}

Hence from Eq. (48), we compute the non-zero $H_{\lambda \nu \mu }$
components as

\begin{equation*}
H_{\theta r\theta }=-H_{r\theta \theta }=\frac{1}{4}r,
\end{equation*}

\begin{equation*}
H_{t\theta r}=-H_{\theta tr}=\frac{1}{8}r^{2}(f^{\theta })^{\prime },
\end{equation*}

\begin{equation*}
H_{tr\theta }=H_{\theta rt}=-H_{r\theta t}=-H_{rt\theta }=\frac{1}{4}%
rf^{\theta }+\frac{1}{8}r^{2}(f^{\theta })^{\prime },
\end{equation*}

\begin{equation}
H_{rtt}=-H_{trt}=\frac{1}{8}f^{\prime }-\frac{1}{4}r(f^{\theta })^{2}-\frac{1%
}{4}r^{2}f^{\theta }(f^{\theta })^{\prime }.  \label{52}
\end{equation}

Subsequently, the spinorial affine connections (47) are evaluated as

\begin{equation}
\Gamma _{t}=\frac{1}{4}\left[ f^{\prime }-r^{2}f^{\theta }(f^{\theta
})^{\prime }\right] \sigma ^{(2)}+i\frac{\sqrt{f}}{4}\left[ 2f^{\theta
}+r(f^{\theta })^{\prime }\right] \sigma ^{(3)},  \label{53}
\end{equation}

\begin{equation}
\Gamma _{r}=\frac{r(f^{\theta })^{\prime }}{4\sqrt{f}}\sigma ^{(1)},
\label{54}
\end{equation}

\begin{equation}
\Gamma _{\theta }=-\frac{r^{2}(f^{\theta })^{\prime }}{4}\sigma ^{(2)}+i%
\frac{\sqrt{f}}{2}\sigma ^{(3)}.  \label{55}
\end{equation}

The UDE (46) can then be explicitly expressed as

\begin{equation}
-i\frac{\sigma ^{(3)}}{\sqrt{f}}\partial _{t}\Psi _{s}+\sqrt{f}\sigma
^{(1)}\partial _{r}\Psi _{s}+\left( \frac{1}{r}\sigma ^{\left( 2\right) }+i%
\frac{f^{\theta }}{\sqrt{f}}\sigma ^{\left( 3\right) }\right) \partial
_{\theta }\Psi _{s}+\left( \frac{f^{\prime }}{4\sqrt{f}}+\frac{\sqrt{f}}{2r}%
\right) \sigma ^{\left( 1\right) }\Psi _{s}-\frac{(f^{\theta })^{\prime }r}{4%
}I_{2X2}\Psi _{s}=\frac{m_{s}}{\hbar }\Psi _{s},  \label{56}
\end{equation}

where $I_{2\times 2}$ is the $2\times 2$ unitary matrix. Equation (56)
matches with the result of \cite{Italy}. Now, using the following ans\"{a}tz
for the spinor:

\begin{equation}
\Psi _{s}=\left\{ 
\begin{array}{c}
\widetilde{A}(t,r,\theta )\exp \left[ \frac{i}{\hbar }I(t,r,\theta )\right]
\\ 
\widetilde{B}(t,r,\theta )\exp \left[ \frac{i}{\hbar }I(t,r,\theta )\right]%
\end{array}%
\right\} ,  \label{57}
\end{equation}

(recall that $I(t,r,\theta )$ represents the action), we obtain a pair of
coupled equations (to the leading order in $\hbar $):

\begin{eqnarray}
\frac{\widetilde{A}}{\sqrt{f}}\partial _{t}I+i\sqrt{f}\widetilde{B}\partial
_{r}I+\left( \frac{\widetilde{B}}{r}-\frac{\widetilde{A}f^{\theta }}{\sqrt{f}%
}\right) \partial _{\theta }I &=&m_{s}\widetilde{A},  \label{58} \\
-\frac{\widetilde{B}}{\sqrt{f}}\partial _{t}I+i\sqrt{f}\widetilde{A}\partial
_{r}I+\left( \frac{\widetilde{B}f^{\theta }}{\sqrt{f}}-\frac{\widetilde{A}}{r%
}\right) \partial _{\theta }I &=&m_{s}\widetilde{B}.  \label{59}
\end{eqnarray}

Equations (58) and (59) have non-trivial solutions for $\widetilde{A}$ and $%
\widetilde{B}$ provided that the determinants of the coefficient matrices
vanish. Hence, we have

\begin{equation}
\frac{1}{f}(\partial _{t}I-f^{\theta }\partial _{\theta }I)^{2}-f(\partial
_{r}I)^{2}-\frac{1}{r^{2}}(\partial _{\theta }I)^{2}-m_{s}^{2}=0.  \label{60}
\end{equation}

By the same process as the previous section, we insert ans\"{a}tz (29) into
Eq. (60) and obtain the following integral solution for $W(r)$:

\begin{equation}
W_{\pm }(r)=\pm \dint \frac{\sqrt{(E+Lf^{\theta })^{2}-f\left[ \left( \frac{L%
}{r}\right) ^{2}+m_{s}^{2}\right] }}{f}.  \label{61}
\end{equation}

The above equation is structurally very similar to Eq. (30). Naturally, Eq.
(61) reduces to Eq. (32) near the event horizon, and consequently, yields
the tunneling rate computed by Eq. (38). We remark that, similar to the
scalar radiation, the temperature of fermions radiated from the event
horizon of a $3D$ RSHBH is the standard $T_{H}$\ (21).

\section{Conclusion}

In this paper, we investigated the HR of scalar and Dirac particles
diverging from the event horizon of a $3D$ RSHBH. For this purpose, we
separated the KGE and UDE on the $3D$ RSHBH geometry using particular ans%
\"{a}tze for the wave functions Eqs. (27) and (57), respectively. We
calculated the quantum tunneling rates of the scalar particles and fermions
using the first-order WKB approximation, thereby demonstrating the effect of
scalar hair on the tunneling rate of a rotating BTZ BH. Remarkably, both
tunneling rates were identical regardless of particle type. After
substituting the tunneling rate into the Boltzmann formula (39), we
recovered the original $T_{H}$\ (21) of the $3D$\ RSHBH.

Finally, whether the results are modified in other hairy BHs, such as BHs
with Abelian Higgs hair \cite{Ana}, is an interesting question, and will be
investigated in our future work.

\end{document}